\documentstyle[preprint,tighten,prc,aps,epsfig]{revtex}

\newcommand{\casc}{$\Xi^-$}
\newcommand{{\kplus}}{K$^+$}
\newcommand{\kminus}{K$^-$}
\newcommand{{\dlam}}{double-$\Lambda$}
\newcommand{\name}[2] {#2$^{\ref{#1}}$}
\newcommand{\dname}[3] {#3$^{\ref{#1},\ref{#2}}$}
\newcommand{\instit}[2]{\refstepcounter{inst}\label{#1}$^{\ref{#1}}$#2\\}

\begin{document}
\draft

\title{Evidence of $\Xi$ hypernuclear production in the
$\mathrm{^{12}C(K^-,K^+)_\Xi^{12} Be}$ reaction}

\author{
\name{inst_cmu} {P.~Khaustov},
\name{inst_bnl} {D.~E.~Alburger},
\name{inst_lanl} {P.~D.~Barnes},
\name{inst_unm} {B.~Bassalleck}, 
\name{inst_cmu} {A.~R.~Berdoz}$^,$\footnotemark[1], 
\name{inst_cmu} {A.~Biglan}$^,$\footnotemark[2],
\name{inst_frei} {T.~B\"{u}rger}, 
\name{inst_cmu} {D.~S.~Carman}$^,$\footnotemark[3],
\name{inst_bnl} {R.~E.~Chrien},  
\dname{inst_manitoba}{inst_triumf}{C.~A.~Davis},
\name{inst_frei} {H.~Fischer}, 
\name{inst_cmu} {G.~B.~Franklin}, 
\name{inst_frei} {J.~Franz},
\name{inst_manitoba} {L.~Gan}$^,$\footnotemark[4],
\name{inst_kyoto} {A.~Ichikawa},
\name{inst_kek} {T.~Iijima},
\name{inst_kyoto} {K.~Imai},
\name{inst_kyoto} {Y.~Kondo},
\name{inst_cmu} {P.~Koran},
\name{inst_manitoba} {M.~Landry},
\name{inst_manitoba} {L.~Lee},
\dname{inst_ub}{inst_unm} {J.~Lowe},
\name{inst_cmu} {R.~Magahiz}$^,$\footnotemark[5], 
\name{inst_bnl} {M.~May},
\name{inst_cmu} {R.~McCrady}$^,$\footnotemark[6], 
\name{inst_cmu} {C.~A.~Meyer},
\name{inst_lanl} {F.~Merrill},
\name{inst_osaka} {T.~Motoba},
\name{inst_manitoba} {S.~A.~Page},
\name{inst_cmu} {K.~Paschke},
\name{inst_bnl} {P.~H.~Pile},
\name{inst_cmu} {B.~Quinn},
\name{inst_manitoba} {W.~D.~Ramsay},
\name{inst_bnl} {A.~Rusek},
\name{inst_bnl} {R.~Sawafta}$^,$\footnotemark[7], 
\name{inst_frei} {H.~Schmitt},
\name{inst_cmu} {R.~A.~Schumacher},
\name{inst_unm} {R.~W.~Stotzer}$^,$\footnotemark[8], 
\name{inst_bnl} {R.~Sutter},
\name{inst_kyoto_s} {F.~Takeutchi},
\name{inst_manitoba} {W.~T.~H.~van~Oers},
\name{inst_kyoto} {K.~Yamamoto},
\name{inst_tsuru} {Y.~Yamamoto},
\name{inst_kyoto} {M.~Yosoi},
\name{inst_kentucky} {V.~J.~Zeps}\\
\vspace{2mm}
(The AGS E885 Collaboration)\\
\vspace{2mm}
}

\newcounter{inst}
\address{
\instit {inst_bnl} {Brookhaven National Laboratory, Upton, NY 11973, USA}
\instit {inst_cmu} {Carnegie Mellon University, Pittsburgh, PA 15213, USA} 
\instit {inst_kek} {KEK, High Energy Accelerator Research Organization,
  Tsukuba, Ibaraki 305, Japan}
\instit {inst_kyoto_s} {Kyoto Sangyo University, Kyoto 603, Japan} 
\instit {inst_kyoto} {Kyoto University, Sakyo-Ku, Kyoto 606, Japan}
\instit {inst_lanl} {Los Alamos National Laboratory, Los Alamos, NM 87545, USA} 
\instit {inst_osaka}
  {Osaka Electro-Communication University, Neyagawa, Osaka 572-8530, Japan}
\instit {inst_triumf} {TRIUMF, Vancouver, BC V6T 2A3, Canada} 
\instit {inst_tsuru} {Tsuru University, Tsuru 402-8555, Japan}
\instit {inst_ub} {University of Birmingham, Birmingham B15 2TT, UK}
\instit {inst_frei} {University of Freiburg, D-79104 Freiburg, Germany} 
\instit {inst_kentucky} {University of Kentucky, Lexington, KY 40507 USA} 
\instit {inst_manitoba}
  {University of Manitoba, Winnipeg, Manitoba R3T 2N2, Canada} 
\instit {inst_unm} {University of New Mexico, Albuquerque, NM 87131, USA} 
}

\footnotetext[1]{
Present address: SFA Inc., Largo, MD 20774, USA 
}
\footnotetext[2]{
Present address: Union Switch \& Signal, Pittsburgh, PA 15219, USA
}
\footnotetext[3]{
Present address:  Ohio University, Athens, OH 45701, USA
}
\footnotetext[4]{
Present address: Hampton University, Hampton, VA 23668, USA
}
\footnotetext[5]{
Present address: Syncsort Inc., Woodcliff Lake, NJ 07675, USA
}
\footnotetext[6]{
Present address: Los Alamos National Laboratory, Los Alamos, NM
87545, USA
}
\footnotetext[7]{
Present address: North Carolina A\&T State University, Greensboro, NC
27411, USA
}
\footnotetext[8]{
Present address: ASAP, Albuquerque, NM 87110, 
USA
}

\date{\today}
\maketitle

\begin{abstract}
The E885 collaboration utilized the 1.8~GeV/c $\mathrm{K^-}$ beam line
at the Alternating Gradient Synchrotron (AGS)
to accumulate greater than
10 times the world's existing data sample of (K$^-$,K$^+$)
events on carbon.  A total of about $\mathrm{3\times10^5}$
(K$^-$,K$^+$) events were collected and analyzed.  $\Xi$ hypernuclear
states are expected to be produced through the reaction $\mathrm{K^- +
^{12}C \rightarrow K^+ + ^{12}_{\;\Xi}Be}.$ A signal could also result
from direct production of $\mathrm{^{11}_{\Lambda}Be + \Lambda}$
without a distinct $\Xi$ intermediate state.  The measured
missing-mass spectrum indicates the existence of a signal below the
threshold for free $\mathrm{\Xi^-}$ production that cannot be
explained by background or effects of limited resolution. Although the
resolution was not sufficient to resolve discrete hypernuclear states,
the excess of events in the region of missing mass, kinematically
inaccessible in free $\Xi^-$ production, can be compared with
theoretical predictions for $\mathrm{^{12}_{\;\Xi}Be}$ production.
Reasonable agreement between the data and theory is achieved by
assuming a $\Xi$-nucleus potential well depth $\mathrm{V_{0\,\Xi}}$ of
about~14~MeV within the Woods-Saxon prescription.
\end{abstract}

\section{Introduction}

With the advent of magnetic spectrometers, it became possible to study
hypernuclei through the reconstruction of missing mass.  Hypernuclei
with one strange quark (S=--1 sector) have long been studied using this
method quite successfully.  Similarly to production of S=--1 hypernuclei in
($\pi^+$,K$^+$) and (K$^-$,$\pi^-$) reactions,
the (K$^-$,K$^+$) reaction can be used to create S=--2 hypernuclei.

Dover and Gal found the $\Xi$-nucleus potential
well depth to be $\approx$~21~MeV -- 24~MeV based on their analysis of
emulsion data~\cite{dover}. The $1s$ and $1p$ single-particle {\casc} states are
expected to be bound for the values of the potential within that range.
Although bound $\Xi$-nucleus states would be broadened by the conversion
$\Xi N\rightarrow\Lambda\Lambda$, calculations have indicated that their
widths could be a few MeV or less~\cite{yamamoto}.
Fukuda {\it et al.} recently reported evidence for bound $\Xi$
hypernuclei produced in the (K$^-$,K$^+$) reaction at
$p_{K^-}$~= 1.6~GeV/c on a scintillating-fiber target (KEK
E224)~\cite{e224}.
Their missing-mass resolution was not
sufficient to resolve discrete hypernuclear states but their data seemed to
favor a $\Xi$-nucleus potential well depth
of $\approx$ 16~MeV.  They concluded that a value of 24~MeV for
the potential was improbable.

The 1.8 GeV/c K$^-$ beam at the AGS was used to create doubly strange
(S=--2) systems in E885 via the $\mathrm{^{12}C(K^-,K^+)X}$ reaction
at $p_{K^-}$~= 1.8~GeV/c.  A high K$^-$ flux and a thick diamond
target~\cite{alburger} combined to
give a sensitivity to S=--2 states of 1.5 counts per nb/sr as compared
to 0.05 counts per nb/sr for E224~\cite{e224}.  Although the
experimental resolution in E885 was somewhat better than the E224
resolution (4.2~MeV/c$^2$ r.m.s. vs. 5.6~MeV/c$^2$ r.m.s.  for
hydrogen kinematics), it too was insufficient to resolve discrete
hypernuclear states. Nevertheless, an estimate of the cross section can be
made; the excitation-energy spectrum can be compared to the
theoretical predictions for different potential well depths folded
with our experimental resolution.~\footnotemark[1]

\footnotetext[1]{Excitation energy E is defined as the
missing mass minus the combined
mass of the $^{11}$B(gs) and the {\casc} (all times c$^2$).  With
this definition the threshold for free {\casc} production is at zero
excitation energy and bound $\Xi$ hypernuclear states have negative
excitation energy.}

The organization of this paper is as follows. In Sec.~\ref{setup} the
experimental setup for E885 is described and the event reconstruction is
briefly discussed. In order to make cross section estimates and
compare the data with theory, the missing-mass spectrum needs to
be normalized.  This normalization procedure
is described in Sec.~\ref{normalization}.  Prior to a comparison with
our data, the theoretical cross section calculation must be folded with
the experimental resolution function. The determination of the
resolution function is described in Sec.~\ref{resolution}.
The theoretical cross sections for a series of
values of the $\Xi$-nucleus potential well depth parameter,
folded with the experimental resolution function, 
are compared with the data in Sec.~\ref{comparison}.
A discussion of background is
presented in Sec.~\ref{background}, followed by our conclusions in
Sec.~\ref{conclusions}.

\section{Experimental setup}
\label{setup}

\begin{figure}
\begin{center}
\epsfig{file=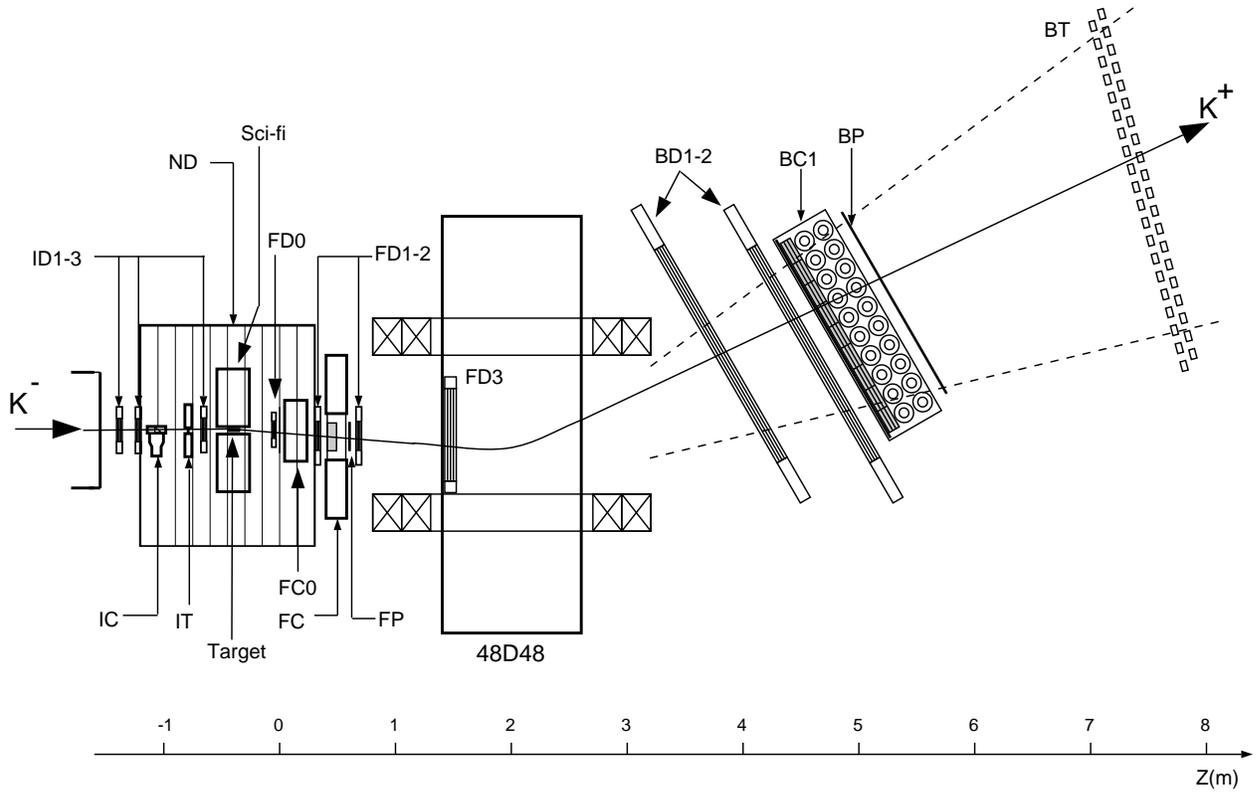,width=17cm,bbllx=80,bblly=250,bburx=520,bbury=550}
\caption{Detector configuration for E885.  The drift chambers ID1-3
determine the incident K$^-$ trajectory and, combined with the beam line hodoscope
MP (not shown here), the K$^-$ momentum.  The drift chambers FD0-3 and BD1-2
determine the K$^+$ trajectory through the 1.4~T 48D48 dipole.
Scintillators IT and BT determine the K$^+$ time-of-flight. Hodoscopes
FP and BP determine the spectrometer acceptance.  Aerogel Cherenkovs
IC, FC, and BC reject pions, while FC0 rejects protons.  Above and
below the target are scintillating fiber detectors. On the left
and right of the target are neutron detectors. The fiber detectors
and neutron detectors were not used in this analysis.}
\label{exp}
\end{center}
\end{figure}

A side view of the E885 detector apparatus is shown in Fig.~\ref{exp}.
The K$^-$
beam was delivered by the AGS D6 beam line\cite{d6} at a rate of
$10^6$ K$^-$/spill with a 1:1 K/$\pi$ ratio. The spill duration and
repetition rate were 1.6 s and 1000 spills/hour, respectively
(44\% duty factor).
The purity and intensity of the kaon beam provided by the AGS
D6 beam line greatly exceeded that available elsewhere.  The beam was
focused in the vertical direction by the last beam line quadrupole
magnet onto a synthetic diamond target, $\mathrm{8\,cm~wide \times
1\,cm~high \times 5\,cm~thick}$ with a density of $\mathrm{3.3\,g/cm^3}$.
Diamond was used as the target
material in order to enhance the {\casc} stopping rate (a property not
important for the direct production discussed in this paper); the
resulting compactness of the target also allowed the application of
tighter geometric cuts.

The momentum of the incident particle was measured using information
from a scintillator hodoscope, MP, located in the beam line at the exit of
the first mass slit (not shown in Fig.~\ref{exp}) and information from
3 drift chambers (ID1-3) located downstream of the last beam line
magnet. The amplitude in the IC Cherenkov counter, along with the
measured particle momentum and the beam line time-of-flight, were used
to suppress triggers from pions.  Additional details regarding the experimental
apparatus can be found in Ref.~\cite{stotzer}.

\begin{figure}
\begin{center}
\epsfig{file=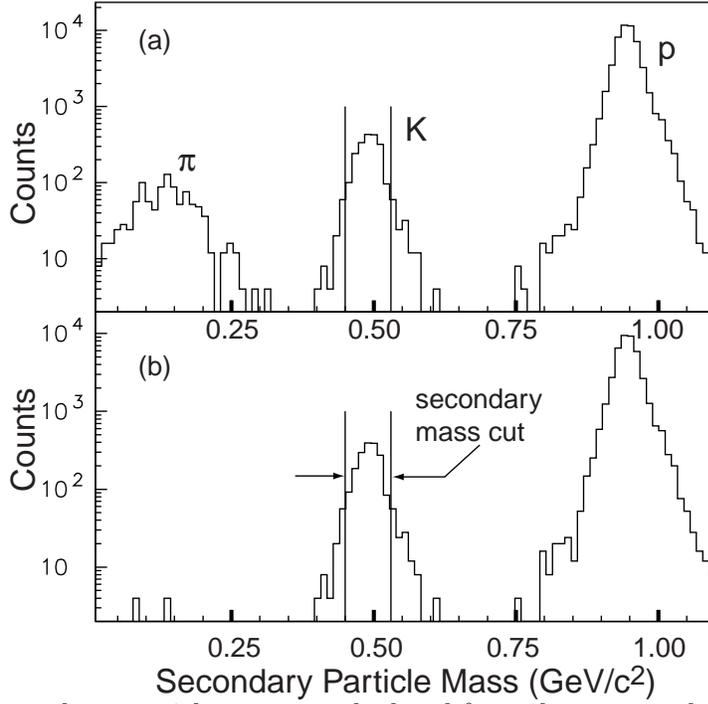,width=10.5cm}
\caption{The secondary particle mass as calculated from the measured
momentum, path length, and time-of-flight through the K$^+$ spectrometer
for a subset of the data.
(a) Events that pass goodness-of-fit, vertex, and multiplicity cuts.
(b) An additional software veto has been applied using the FC pulse
height. Events in (b) that fall within the secondary-mass cut were
used in the analysis.} 
\label{secm}
\end{center}
\end{figure}

The momentum of the outgoing K$^+$ was measured in the 1.4~T 48D48 magnetic
spectrometer. The amplitude in the FC Cherenkov counter, along with the
measured momentum and IT-BT time-of-flight, was used to select kaons. In
addition to pions, protons were another source of background in the
spectrometer.  To reduce the hardware trigger rate, an aerogel
Cherenkov counter, FC0, was used to suppress protons; its refraction
index was selected to be sensitive to kaons but not protons from inelastic
events.  The
measured mass spectrum for a sample of secondary particles 
is shown in Fig.~\ref{secm}.  This shows that the kaons were
cleanly separated from the pions and protons that passed the hardware
triggers after goodness-of-fit and vertex cuts were applied. The pion
peak in Fig.~\ref{secm}a is due to leakage of the hardware FC pion
veto.  These events were virtually eliminated by an additional
software cut on FC pulse height.  The results are shown in
Fig.~\ref{secm}b.  The measured momenta of the incident and secondary
particles allow the calculation of the missing mass.

\section{Normalization}
\label{normalization}

Our data were normalized to the known forward-angle
cross section of free {\casc}
production on hydrogen (contained in a CH$_2$ target).  Due to the low
statistics in the bound region, only angle-averaged differential cross
sections could be obtained.  These angle-averaged cross sections are denoted
here as $\displaystyle \left\langle \frac{d^2\sigma}{d\Omega dE}
\right\rangle$.
We present results averaged over
$\mathrm 0^\circ<\theta_{K^+}<14^\circ$, corresponding to the full acceptance
of the spectrometer, and for $\mathrm 0^\circ<\theta_{K^+}<8^\circ$.

The normalization effectively scaled the diamond target results to
agree with the {\casc} production off carbon in a CH$_2$ target, which
was, in turn, normalized to the $\Xi$ production off hydrogen in the
CH$_2$ target.  This procedure corrected for differences in the event
reconstruction efficiency between the low-intensity CH$_2$ running and
the high intensity diamond target running. These differences
were brought about, in part, by tight multiplicity cuts on detector MP.

The  scale for the excitation-energy spectrum was set using the
equation:
\begin{eqnarray} 
&& \left\langle \frac{d^2\sigma}{d\Omega dE} \right\rangle^C = 
 A \left( \frac{dN}{dE} \right)_{diam}^C,
\label{crap3}
\end{eqnarray}
where the normalization factor is given by

\[ A = \frac{\alpha^H}{\alpha^C(E)} \, \frac{\lambda_{CH_2}^H}{\lambda_{CH_2}^C}
\, \frac{N_{CH_2}^C}{N_{diam}^C N_{CH_2}^H}
\, \left\langle \frac{d\sigma}{d\Omega} \right\rangle^H. \]

The following definitions were used:
\begin{itemize}

\item $\displaystyle\left\langle \frac{d^2\sigma}{d\Omega dE} \right\rangle^C$  -- 
angle-averaged
double-differential cross section for hypernuclear production on carbon.

\item $\displaystyle\left( \frac{dN}{dE} \right)_{diam}^C$ -- 
number of hypernuclear-production events in an excitation energy interval~$dE$
from the diamond target.

\item $\displaystyle\alpha^H$ -- angle- and energy-averaged 
acceptance for free $\Xi^-$ production on hydrogen.

\item $\displaystyle\alpha^C(E)$ -- angle-averaged
acceptance for hypernuclear production on carbon. The argument $E$ indicates the
dependence on excitation energy.

\item $\displaystyle\lambda_{CH_2}^H$ -- target thickness for hydrogen 
in CH$_2$ target.

\item $\displaystyle\lambda_{CH_2}^C$ -- target thickness for carbon
in CH$_2$ target.

\item $\displaystyle N_{CH_2}^C$ -- number of events of free $\Xi^-$ production on 
carbon in CH$_2$ target.

\item $\displaystyle N_{CH_2}^H$ -- number of events of free $\Xi^-$ production on 
hydrogen in CH$_2$ target.

\item $\displaystyle N_{diam}^C$ -- number of events of free $\Xi^-$ production on 
carbon in diamond target.

\item $\displaystyle\left\langle \frac{d\sigma}{d\Omega} \right\rangle^H$ --
angle-averaged differential cross section of $\Xi^-$ production on hydrogen.

\end{itemize}

Since the spectrometer acceptance for hypernuclear production
varies very little with
energy over the small region of interest about zero excitation energy,
instead of the energy-dependent acceptance $\alpha^C(E)$, the
acceptance was calculated for a fixed excitation energy of $-10$~MeV.
The angle-dependent acceptance used to determine
$\alpha^C(E)$ was generated with a Monte Carlo simulation and is shown
as the solid line in Fig.~\ref{angular}.

A summary of the parameters used for normalization in Eq. (\ref{crap3}) is
given in Table~\ref{table1} for two different acceptances; the first
one is full spectrometer acceptance ($\theta_{K^+}~<~14^\circ$) and
the second is for limited acceptance
($\theta_{K^+}~<~8^\circ$)~\footnotemark[2].
\footnotetext[2]{Note that in both cases we use the value of
35~$\mu$b/sr for the elementary production cross section which is
compatible with the measured value in Ref.~\cite{iijima}.  This factor
should be divided out when comparing these results to a calculation of
the effective proton number.}
The following values for the normalization factor $A$, giving the
angle-averaged differential cross section per event were obtained:
\[
A = \left\{ 
\begin{array}{l l}
0.64 \pm 0.05\; \mathrm{nb/sr} & \mathrm{for\;\theta_{K^+}~<  14^\circ} \\
2.14 \pm 0.20\; \mathrm{nb/sr}&  \mathrm{for\;\theta_{K^+}~<  8^\circ}. \\
\end{array} 
\right.
\label{normfactor}
\]
This factor relates the counts per MeV to the angle-averaged
differential cross section per MeV. The statistical errors given for
the normalizing factor $A$ were due mostly to the limited amount of
calibration data, namely the number of free {\casc} production events
on hydrogen and carbon in the CH$_2$ target.  The systematic
uncertainty, excluding the uncertainty of the elementary cross section
of {\casc} production on hydrogen, is expected to be below the
statistical error. The uncertainty of the elementary cross section for
{\casc} production on hydrogen does not affect the comparison between
data and theory because the theoretical DWIA calculations are
normalized to the same elementary cross section. Thus the
data-to-theory ratio is insensitive to the assumed elementary
cross-section.

\section{Missing-mass resolution}
\label{resolution}

Data sets were taken with two different CH$_2$ targets and were used
to study the missing-mass resolution.  The first set used an 8~cm long
target and the second set, which was acquired over a shorter time
period, used a 13~cm long target.  The first data set was used to
study the spectrometer's intrinsic resolution and, in particular,
to evaluate the tails of the resolution function. The second set was used
to derive the width of the resolution function.  The contribution of
energy-loss effects to the resolution for the 13~cm long CH$_2$ target
is expected to be similar to the contribution from the 5~cm long
diamond target.

\begin{figure}
\begin{center}
\epsfig{file=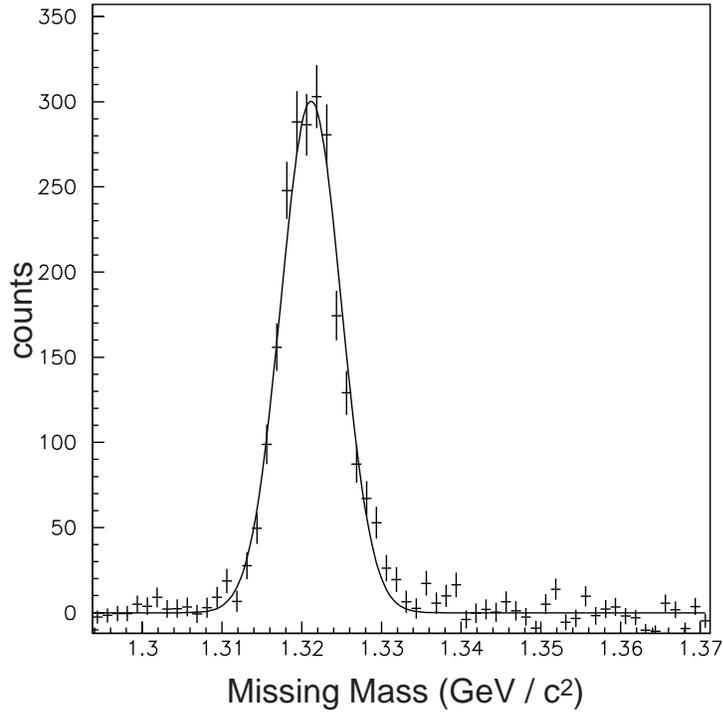,width=10.5cm}
\caption{Missing mass for $\Xi^-$ production on hydrogen in the 8~cm long 
CH$_2$ target. Background from reactions on carbon has been subtracted using
a data sample, properly normalized, taken with diamond target.
The solid line shows the best Gaussian fit.}
\label{sub}
\end{center}
\end{figure}

Due to the two-body nature of {\casc} production on hydrogen, the
missing mass spectrum is a peak, whose width equals the experimental
resolution, on top of a background from {\casc} production on
carbon. In addition to the peak width, we are also interested in the
tail of the resolution. The study of the resolution tails is important;
the small hypernuclear-production signal is not well separated in missing mass
from the large signal of free {\casc} production. The background under the
hydrogen peak,
consisting of {\casc} production on carbon, is subtracted using the
missing-mass spectrum of the carbon target. The resulting
background-subtracted spectrum is shown in Fig.~\ref{sub}.
The width of the peak
from a Gaussian fit is 3.8~MeV/c$^2$ r.m.s.. It can be seen from the
plot that the resolution is reasonably close to being Gaussian.  The
missing mass resolution for the 13~cm CH$_2$ target was measured to be
4.2~MeV/c$^2$ r.m.s. for hydrogen kinematics.

The mass resolution is a function of the target mass.  The resolution
specific to {\casc} production on hydrogen (which we measure directly)
was used to infer the missing-mass resolution for carbon kinematics
using simple kinematical considerations.  A factor of 1.45 relates the
missing-mass resolutions in carbon and hydrogen kinematics.  This
value results in a missing-mass resolution for carbon kinematics of
4.2~MeV/c$^2\times$1.45 = 6.1~MeV/c$^2$. The peak of {\casc}
production on hydrogen was also used for the energy-scale
calibration. The position of the peak was adjusted to be at the
{\casc} mass value and the accuracy of the energy-scale calibration
for the diamond target was estimated to be better than 0.5~MeV.

\section{Comparison of the data and theory}
\label{comparison}

The theoretical (K$^-$,K$^+$) double differential cross section 
for $\mathrm{^{12}_{\;\Xi}Be}$ production, 
$ \frac{d^2\sigma(\theta )}{d\Omega_L dE_\Xi}$, was calculated, 
for both bound and unbound {\casc}s, in the Kapur-Peirls framework 
which was developed originally for an estimate of the $(\pi^+,K^+)$ 
strength function~\cite{motoba}. The DWIA calculation was performed 
for a K$^-$ momentum of 1.8~GeV/c by assuming the K-N 
elementary cross sections: $\sigma(K^-N)$=2.90 fm$^2$ and 
$\sigma(K^+N)$=1.99 fm$^2$. Further details of the calculation 
can be found in the recent references~\cite{yamamoto} and \cite{ikeda} 
in which the (K$^-$,K$^+$) cross sections for the $^{12}$C and 
$^{16}$O targets have been estimated for $p_{K^-}=1.6$~GeV/c. 

The calculation was performed for $0^{\circ}<\theta_{K^+}<16^{\circ}$ 
and a series of Woods-Saxon {\casc} well depth parameter values: 
V$_{0\Xi}=$ 12, 14, 16, 18, and 20 MeV. 
The radius and skin depth of the $\Xi$ potential
were fixed at $R=1.1A^{1/3}$~fm and
$a_\Xi$=0.65 fm.  The proton wave function in the $\mathrm^{12}C$ target was
generated using a Woods-Saxon potential with
V$_{0\,N}$ = 50~MeV, $R=1.1A^{1/3}$~fm and $a_N$=0.65~fm.
The elementary {\casc} production cross section was set to 35$\mu$b/sr
to be compatible with the normalization of our experimental data.
The kinematic factor $\alpha$, which accounts for the transformation between
the two-body and A-body frames, was set to 0.73.

\begin{figure}
\begin{center}
\epsfig{file=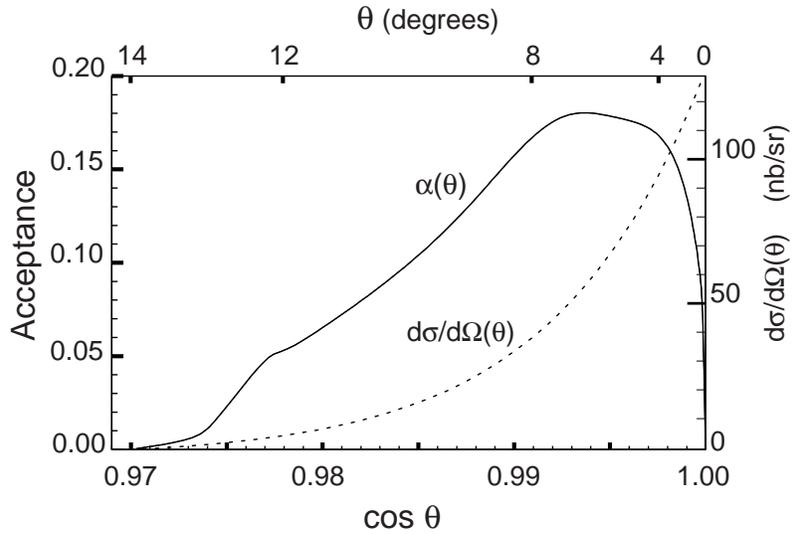,width=10.5cm}
\caption{The angular acceptance, $\alpha$($\theta$),
generated by a Monte Carlo simulation is
shown as the solid line.  The dashed line shows the results of the DWIA
calculation of the angular dependence of the cross section of the ground
state using a potential depth of V$_{0\,\Xi}$=14~MeV.}
\label{angular}
\end{center}
\end{figure}

\begin{figure}
\begin{center}
\epsfig{file=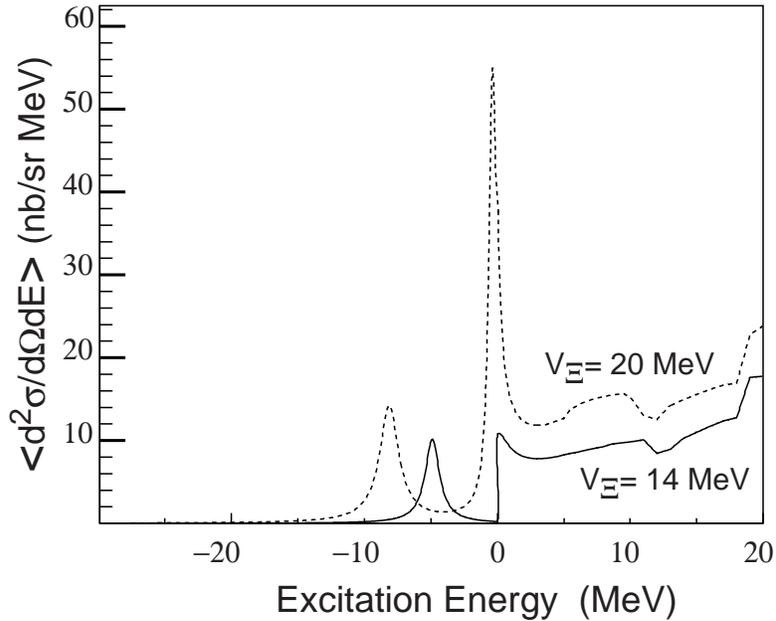,width=10.5cm}
\caption{Results of DWIA calculations, before folding by the
experimental energy resolution, for the
$\mathrm{^{12}C(K^-,K^+)_\Xi^{12} Be}$ reaction for
V$_{0\,\Xi}$=14~MeV and 20~MeV.  The cross section has been averaged
over the kaon angular range $\mathrm0<\theta_{K^+}<14^\circ$.}
\label{theory}
\end{center}
\end{figure}

As a typical example of the angular dependence of the differential cross 
section, the dashed line in Fig.~\ref{angular} shows the 
case of the ground state for V$_{0\,\Xi}$=14 MeV.
In order to make a comparison between the theory and the 
data, we calculate the angle-averaged differential cross sections, 
$\displaystyle \left\langle\frac{d^2\sigma}{d\Omega dE}(E) 
\right\rangle $ and fold the results with the experimental resolution.
However, we first present theoretical 
$^{12}$C(K$^-$,K$^+)^{12}_{\Xi }$Be spectra which have not been 
folded by the experimental energy resolution but have been 
angle-averaged over $0^{\circ} <\theta_{K^+}<14^{\circ}$. As shown 
in Fig.~\ref{theory}, the result for V$_{0\Xi}$=20 MeV (dashed line)
has two bound-state peaks, corresponding to the $\Xi^-$ $s$- and 
$p$-orbitals. The widths of these peaks are determined
using a one-boson-exchange model to estimate the rate for
the $\Xi N\rightarrow \Lambda\Lambda $ conversion.
For the case of V$_{0\Xi}$=14 MeV, the $p$ state
is not bound but it is calculated
as a resonance state in the continuum; therefore a sudden rise 
is seen just above the threshold in Fig.~\ref{theory}.

\begin{figure}
\begin{center}
\epsfig{file=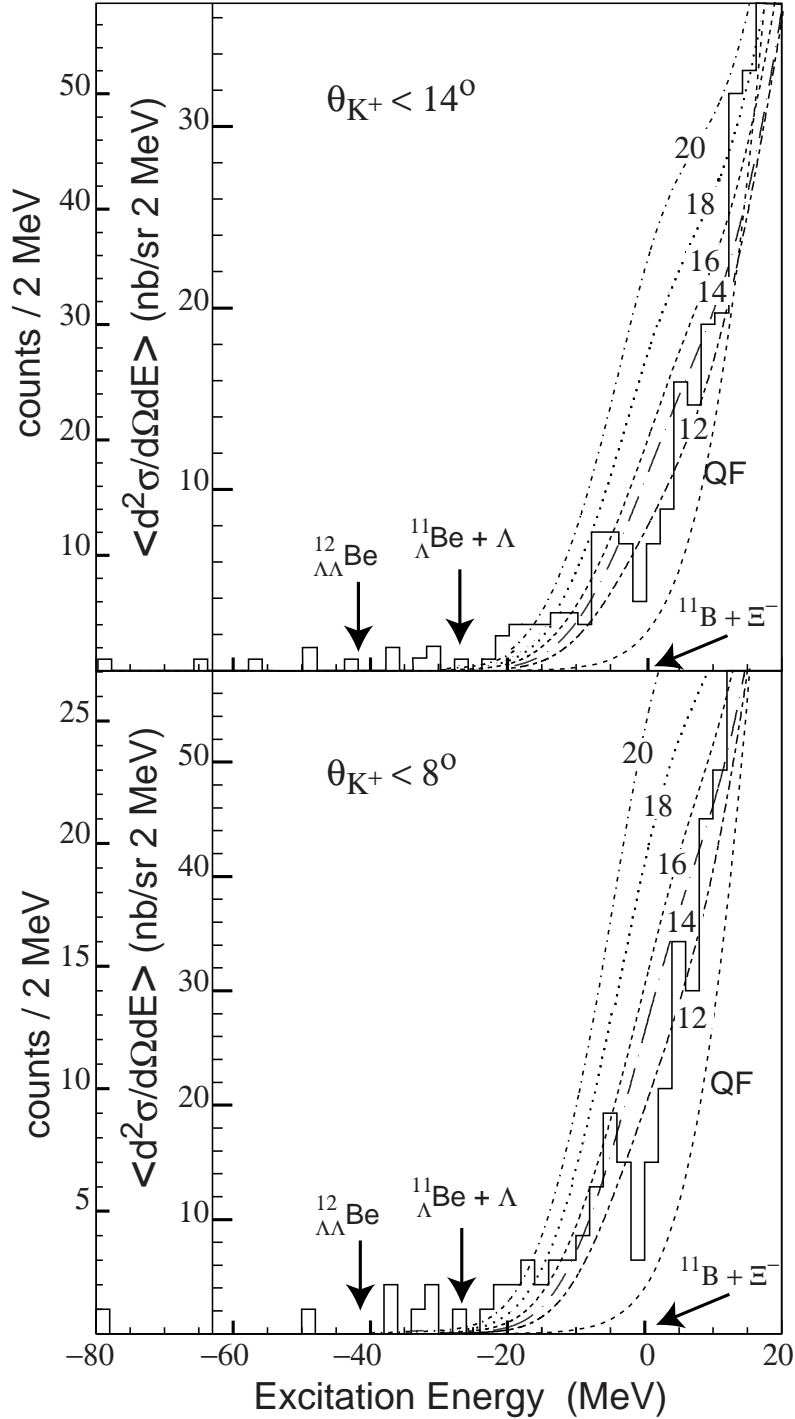,width=10.5cm}
\caption{Excitation-energy spectra from E885 for $^{12}$C({\kminus},{\kplus})X
for $\theta_{K^+} <$14$^\circ$ (top figure) and $\theta_{K^+} <$8$^\circ$
(bottom figure) along with $\mathrm{^{12}_{\;\Xi}Be}$ production
theoretical curves for V$_{0\,\Xi}$ equal to 20, 18, 16, 14, and 12~MeV.
The results of a quasi-free $\Xi$ production
calculation are also shown (curve QF).
The expected location of the ground
state of $^{12}_{\Lambda\Lambda}$Be and the thresholds for
\mbox{$^{11}_\Lambda$Be + $\Lambda$} and
\mbox{$\mathrm{^{11}B +
\Xi^-}$} production are indicated with arrows.}
\label{exc_spec} 
\end{center}
\end{figure}

Figure~\ref{exc_spec} shows experimental excitation 
energy histograms for $^{12}$C({\kminus},{\kplus})X 
for two different limits on the scattering angle of the outgoing 
{\kplus}, $\theta_{K^+}<14^\circ$ and $\theta_{K^+}<8^\circ$.
The data clearly show an enhancement around zero
excitation energy when compared to a Monte Carlo simulation based on
quasi-free $\Xi$ production which has been normalized to the total
number of $^{12}$C({\kminus},{\kplus})X events (curve QF).

In the same figure, the $\mathrm{^{12}_{\;\Xi}Be}$ production 
theoretical curves for several $\Xi$ potential well depths, folded 
with the 6.1~MeV~r.m.s. experimental resolution, are shown
for comparison with the data. The expected location of the ground
state of $^{12}_{\Lambda\Lambda}$Be (assuming a total binding energy
of the $\Lambda$'s, $\mathrm{B_{\Lambda\Lambda}}$, of 25~MeV) and the
thresholds for \mbox{$\mathrm{^{11}_\Lambda Be + \Lambda}$} and
\mbox{$\mathrm{^{11}B + \Xi^-}$} production are indicated.
The normalization calculation for the case $\theta_{K^+}<8^\circ$ is
less sensitive to the model of angular dependence because the
spectrometer acceptance is fairly flat over this region but drops
rapidly for $\theta_{K^+}>8^\circ$ as shown in Fig.~\ref{angular}; we
present the results for both the entire acceptance and for
$\theta_{K^+}<8^\circ$ in Fig.~\ref{exc_spec}.

Visual inspection shows that the theoretical curve for the value of
the $\Xi$-nucleus potential well depth V$_{0\,\Xi}$~=~14~MeV agrees
with the data reasonably well in the region of excitation energy
\mbox{-20~MeV$<$E$<$0~MeV} and much better than the curve for
V$_{0\,\Xi}$~=~20~MeV. If any of the observed signal results from
direct two-$\Lambda$ production without an
intermediate $\mathrm{^{12}_\Xi Be}$ state, the discrepancy between
the V$_{0\,\Xi}$~=~20~MeV results
and the remaining experimental signal becomes even
larger.
The comparison of data to theory in the bound region can be quantified by
considering the results for the angle-averaged cross section
integrated over the \mbox{-20~MeV$<$E$<$0~MeV} excitation region.  The
results are presented in Table~\ref{table2}.  It can be seen that the
DWIA calculation predicts a yield more than a factor of two larger
than the measured cross section in this region when a $\Xi$-nucleus
potential well depth of 20~MeV is used.

\section{Background}
\label{background}

Events due to $\Xi$ production off the hydrogen contained in the target and
target-area scintillators produce a small peak at $\approx$+90 MeV
in excitation energy; this peak represents only a few per cent of the
total events and thus this contamination does not affect the analysis
near zero excitation energy.
Background in the region of negative values of excitation energy can be
divided into two groups:
\begin{itemize}

\item Flat background caused by particle misidentification, severe
tracking errors, etc. Such background might have some structure over a
wider region of missing mass, but on a small, $\approx$~100~MeV scale,
it should be flat. This was confirmed when observing the background in
the bound region obtained by using a loose set of cuts (hence
considerable background contamination).

\item Leakage from E$>$0 due to finite missing-mass resolution.

\end{itemize}
Background of the first kind can be easily estimated by observing the
region further below zero where leakage becomes negligible.  There are
6 events in the interval of excitation energy [-80~MeV,-40~MeV] for
$\theta_{K^+}~<~14^\circ$ (see Fig.~\ref{exc_spec}).
Since the flatness of the background was
confirmed by loosening the cuts, an extrapolation can be made that the
number of background events of the first kind in the excitation-energy
region [-20~MeV,0~MeV] is about 3. It can be concluded then that the
background of the first kind contributes only a small fraction of the
total signal in the $\Xi$ hypernuclear production region, and it is
ignored when comparing the data to theory.

Background of the second kind is taken into account when the
theoretical cross section is folded with the experimental resolution.
The number of events in the E$<$0 region that leak from the E$>$0
region depends on the shape and magnitude of the excitation-energy
spectrum for E$>$0 and on the resolution function shape.  We found
that the signal observed in the bound region cannot be explained by
leakage from the E$>$0 region. This is already evidenced by the good
agreement between the data and theory for V$_{0\,\Xi}$~=~14~MeV in the
E$<$0 region, since non-zero cross section of hypernuclear production
is essential to explain the events in the bound region. We, however,
made two additional checks to demonstrate it.  First, a Monte Carlo
excitation-energy spectrum, where only free {\casc} production is
simulated and no events in the E$<$0 region are produced, was folded
with the measured experimental resolution.  It was found that only
$\approx$ 10 events are expected to have E$<$0 for $\theta_{K^+}~<~14^\circ$
(the observed number is 67).
Second, a different resolution function was tried with very
large tails added to the basic Gaussian resolution function. Even with
those large tails, which are clearly ruled out by the resolution
measurement, the signal in the bound region due to leakage was still a
few times below the observed level.

\section{Conclusions}
\label{conclusions}

The measurement of the missing mass in the (K$^-$,K$^+$) reaction on
carbon in E885 allowed the determination of the excitation-energy
spectrum of directly produced S=--2 systems.  The signal in the
missing-mass region below the threshold for free {\casc} production was
examined in order to extract information on the existence of $\Xi$
hypernuclei and the strength of the $\Xi$-nucleus potential.  Our analysis
has shown that the events in the bound region of the missing-mass
spectrum could not be explained by non-S=-2 background or effects of limited
resolution, and these events are consistent with hypernuclear
production. Possible sources of the signal in the bound region
are the production of one or two $\Lambda$s in the continuum or 
double-$\Lambda$ hyperfragments. The threshold for
$^{11}_\Lambda$Be + $\Lambda$
is estimated to be at E~$\approx$~-27~MeV. This channel may dominate
the other two-$\Lambda$ channels since the two-step
(K$^-$,$\pi$),($\pi$,K$^+$) process involves one low-momentum $\Lambda$
and one high-momentum $\Lambda$.  Its cross section is expected to be several
orders of magnitude larger than the production cross section for
double-$\Lambda$ hypernuclei\cite{dover80,chrien}.  No quantitative estimate
of the contribution of these processes in this kinematic regime has been made
and final interpretation of the excitation spectrum in terms of one-step
$\Xi$-hypernuclear production verses two-$\Lambda$ production without an
intermediate $\Xi$-state requires additional theoretical work.

A comparison of the data and the theoretical calculations for the cross
sections of $\Xi$ hypernuclear production, folded with the
experimental resolution, shows a reasonable agreement of the data and
theory for V$_{0\,\Xi}$~=~14~MeV in the region of excitation energy
-20~MeV$<$E$<$0~MeV and a significant disagreement between the data
and theory for V$_{0\,\Xi}$~=~20~MeV. These results are consistent with
the conclusions of Ref.~\cite{e224}.  If a significant portion of the
events in this region correspond to two-$\Lambda$ production
without an intermediate
$\Xi$ hypernucleus, the discrepancy between the calculation using this
well depth and our results would be even larger.  We conclude that the
results are consistent with the theoretical predictions when
a potential depth of 14~MeV or less is assumed.

The accuracy of DWIA, which was used for the cross section
calculation, was nominally assumed to be 30\% in Ref~\cite{e224}.  
Although the DWIA framework has been successful in describing the production
of single-$\Lambda$ hypernuclei using the ($\pi^+$,K$^+$) and (K$^-$,$\pi^-$)
reactions, it is possible that
that the DWIA accuracy for the (K$^-$,K$^+$) reaction is
worse due to the higher momentum transfer. Therefore a value
for the $\Xi$-nucleus potential well depth as high as 20~MeV, though
less favored, cannot be completely ruled out.  Additional theoretical
calculations of the complete $\mathrm{^{12}C(K^-,K^+)X}$ excitation
spectrum are needed to determine the parameter space that is compatible
with our results.

This work is supported in part by the U.~S. Department of Energy under
contracts DE-FG02-91ER40609, DE-AC02-76H00016, and DE-FG03-94ER40821,
by the German Federal Minister for Research and Technology (BMFT)
under contract No.  06 FR 652, by the United Kingdom SERC,
by the Natural Sciences and Engineering Research Council of Canada,
and by the Japanese Society for the Promotion of Science.

\newpage
\mediumtext
\begin{table}
\centering \footnotesize \setlength{\tabcolsep}{0.15cm}
\begin{center}
\begin{tabular}{ccccccc}
maximum $\theta_{K^+}$ & \raisebox{-0.9ex}{$\alpha ^H$} & 
\raisebox{-0.99ex}{$\alpha ^C(-10~MeV)$}  & 
\raisebox{-0.99ex}{$\lambda_{CH_2}^H/\lambda_{CH_2}^C$} & 
\raisebox{-0.99ex}{$N^C_{CH_2}/N^C_{diam}$} & 
$\langle d\sigma / d\Omega \rangle^H$  & 
\raisebox{-0.99ex}{$N_{CH_2}^H$} \\
(degrees) &  &  &  &   & ($\mathrm{\mu b/sr}$)  &   \\
\tableline
14 & 8.1\% & 14.8\% & 2 & 1/125.6 & 35 & 480 \\
8 & 14.4\% & 15.3\% & 2 & 1/127.9 & 35 & 240
\end{tabular}
\caption{A summary of parameters used for normalization. The value
used for the cross section of {\casc} production on hydrogen is
compatible with the value measured by T.~Iijima {\it et
al.}~\protect\cite{iijima}.}
\label{table1}
\end{center}
\end{table}

\mediumtext
\begin{table}
\begin{center}
\begin{tabular}{lrrr}
 & \multicolumn{2}{c}{This Work} & {KEK E224} \\
 & $\theta_{K^+}<$8$^\circ$ & $\theta_{K^+}<$14$^\circ$  \\
\tableline
Counts                              & 42 & 67 &  3  \\
Sensitivity  (counts per nb/sr) & 0.47 & 1.6 & 0.05 \\
$\langle d\sigma / d\Omega \rangle$ Experiment (nb/sr)
                 & 89$\pm$14 & 42$\pm$5 & 60$\pm$45 \\
$\langle d\sigma / d\Omega \rangle$ V$_{0\,\Xi}$=12~MeV (nb/sr)
                 & 67 & 26 & - \\
$\langle d\sigma / d\Omega \rangle$ V$_{0\,\Xi}$=14~MeV (nb/sr)
                 & 92 & 37 & - \\
$\langle d\sigma / d\Omega \rangle$ V$_{0\,\Xi}$=16~MeV (nb/sr)
                 & 115 & 46 & 75 \\
$\langle d\sigma / d\Omega \rangle$ V$_{0\,\Xi}$=18~MeV (nb/sr)
                 & 166 & 68 & - \\
$\langle d\sigma / d\Omega \rangle$ V$_{0\,\Xi}$=20~MeV (nb/sr)
                 & 226 & 93 &170
\end{tabular}
\caption{The experimental results integrated over the excitation
energy \mbox{-20~MeV$<$E$<$0~MeV} are shown.
The theoretical values for the averaged differential cross sections,
after convoluting with the experimental resolutions and integrating
over the same energy range, are shown as a function of
the $\Xi$-nucleus potential well depth parameter, V$_{0\,\Xi}$.
The corresponding values for the KEK E224 experiment~\protect\cite{e224},
p$_{K^-}$=1.66~GeV/c, are shown where available.}
\label{table2}
\end{center}
\end{table}

\widetext
\newpage

\end{document}